\documentclass[aps,prd,onecolumn,groupedaddress,showpacs,nofootinbib,amssymb]{revtex4}
\usepackage[dvips]{graphicx,color}
\usepackage{amssymb}
\usepackage{amsmath}
\usepackage{graphicx}
\usepackage{amsfonts}

\newcommand{\be}{\begin{equation}}
\newcommand{\ee}{\end{equation}}
\newcommand{\bea}{\begin{eqnarray}}
\newcommand{\eea}{\end{eqnarray}}
\newcommand{\beaa}{\begin{eqnarray*}}
\newcommand{\eeaa}{\end{eqnarray*}}

\newcommand{\nn}{\nonumber \\}
\newcommand{\e}{\mathrm{e}}

\begin{document}

\title{Viable Mimetic Completion of Unified Inflation-Dark Energy Evolution in Modified Gravity}
\author{S.~Nojiri,$^{1,2}$\,\thanks{nojiri@gravity.phys.nagoya-u.ac.jp}
S.~D.~Odintsov,$^{3,4}$\,\thanks{odintsov@ieec.uab.es}
V.~K.~Oikonomou,$^{5,6}$\,\thanks{v.k.oikonomou1979@gmail.com}}
\affiliation{$^{1)}$ Department of Physics, Nagoya University, Nagoya 464-8602,
Japan \\
$^{2)}$ Kobayashi-Maskawa Institute for the Origin of Particles and the
Universe, Nagoya University, Nagoya 464-8602, Japan \\
$^{3)}$Institute of Space Sciences (IEEC-CSIC)\\
C. Can Magrans s/n, 08193 Barcelona, SPAIN\\
$^{4)}$ ICREA, Passeig Luis Companys, 23,
08010 Barcelona, Spain\\
$^{5)}$ Tomsk State Pedagogical University, 634061 Tomsk, Russia\\
$^{6)}$ Laboratory for Theoretical Cosmology, Tomsk State University of Control Systems
and Radioelectronics (TUSUR), 634050 Tomsk, Russia\\
}

\begin{abstract}
In this paper, we demonstrate that a unified description of early and late-time acceleration is possible in the context of mimetic $F(R)$ gravity. We study the inflationary era in detail and demonstrate that it can be realized even in mimetic $F(R)$ gravity where traditional $F(R)$ gravity fails to describe the inflation. By using standard methods we calculated the spectral index of primordial curvature perturbations and the scalar-to-tensor ratio. We use two $F(R)$ gravity models and as it turns out, for both the models under study the observational indices are compatible with both the latest Planck and the BICEP2/Keck array data. However, this is only true under some model-dependent fine-tuning, which constrains the models we study. Finally, the graceful exit from inflation issue is addressed, and as we show, the curvature perturbations may trigger the graceful exit from inflation when the slow-roll era ends. However, fine-tuning is needed in order to produce enough inflation by the end of the slow-roll era.
\end{abstract}

%PACS numbers: 04.50.Kd, 95.36.+x, 98.80.-k, 98.80.Cq

\maketitle

\section{Introduction}

Modified gravity offers a consistent theoretical framework for the successful description of the Universe evolution. After the striking observation of the late-time acceleration in the late 90's \cite{Riess:1998cb}, many different forms of modified gravity have been proposed, for example see the reviews
\cite{Nojiri:2006ri,Capozziello:2010zz,Avelino:2016lpj,Nojiri:2010wj,Capozziello:2011et,delaCruzDombriz:2012xy}. One of the most difficult aims is to find a unified description of early and late-time acceleration, with the $F(R)$ theories being the most promising theories of this kind \cite{Nojiri:2003ft,Appleby:2007vb,Nojiri:2007as,Nojiri:2007cq,Cognola:2007zu}. A recent elegant proposal for a modified gravity theory was introduced in Ref.~\cite{Chamseddine:2013kea} and it was called mimetic gravity, with the mimetic terminology referring to the fact that dark matter is mimicked by a geometrical term which originates from the conformal degree of freedom of the metric. It is exactly because of this that mimetic gravity is an elegant modified gravity theory since no new term is added by hand and one actually exploits the conformal internal degree of freedom. Later on the mimetic gravity was further studied in
Refs.~\cite{Chamseddine:2014vna,Hammer:2015pcx,Golovnev:2013jxa,Momeni:2014qta,Matsumoto:2015wja,Leon:2014yua,Haghani:2015zga,Myrzakulov:2015kda,Cognola:2016gjy,Nojiri:2014zqa,Astashenok:2015haa,Raza:2015kha,Myrzakulov:2015qaa,Odintsov:2016imq,Odintsov:2015wwp}, and also was further generalized in the context of $F(R)$ gravity in Ref.~\cite{Nojiri:2014zqa}. Particularly, in Ref.~\cite{Nojiri:2014zqa}, the Lagrange multiplier formalism was used \cite{Lim:2010yk,Capozziello:2010uv}, and the resulting theory has three vital components, the $F(R)$ gravity, the potential and the Lagrange multiplier. In effect, these components of the theory offer much freedom for the realization of various cosmological scenarios, see for example
\cite{Odintsov:2015wwp}. Also, mimetic $F(R)$ gravity may not only unify inflation with the dark energy era, but naturally introduces gravitational dark matter.

The purpose of this paper is two fold. Firstly, we will demonstrate how a unified description of early and late-time acceleration can be achieved in the context of mimetic $F(R)$ gravity, and secondly we shall investigate how a viable inflationary era can be generated by a mimetic $F(R)$ theory. Particularly, the viability of each inflationary scenario under study is quantified in terms of the inflationary observational indices, the spectral index of primordial curvature perturbations $n_s$ and the scalar-to-tensor ratio $r$. Using standard methods
\cite{Noh:2001ia,Hwang:2001qk,Noh:2004rt,Hwang:2001pu} we shall calculate the observational indices and we shall investigate when the resulting cosmologies are compatible with the latest Planck
\cite{Ade:2015lrj,Planck:2013jfk} and BICEP2/Keck Array data \cite{Array:2015xqh}. For our study we shall use two quite well known $F(R)$ gravities, the $R^2$ model and a power-law model of the form $F(R)=\alpha R^n$. The $R^2$ model is already compatible with the Planck data, and as we demonstrate the mimetic model is also in agreement with the Planck and BICEP2/Keck Array data. More interestingly, the power-law model in the context of non-mimetic $F(R)$ gravity yields a scalar-to-tensor ratio which is not compatible with the observational data. However, in the mimetic version of the theory, the power-law model is in agreement with all the latest observational data, but at the expense of fine-tuning on the free parameters of the theory. Thus the mimetic theory has some appealing features-apart from the fine-tuning-that need to be taken seriously into account. One could claim that in this theory many cosmological scenarios can be realized and nothing new is predicted from the theory. In our opinion, the mimetic theory is ``economic'' in the sense that nothing is actually added by hand and only the conformal degree of freedom of the metric is used. Therefore, the theory deserves some attention and many theoretical implications resulting from it should be studied before the theory can be considered as a mathematical construction only.

This paper is organized as follows: In section II we present in brief the essential features of mimetic $F(R)$ gravity and also we discuss how the Newton law for gravity is affected by the mimetic potential and the Lagrange multiplier. Also we investigate how the unification of late and early-time acceleration can be achieved in the context of mimetic $F(R)$ gravity. In section III we demonstrate how various viable inflationary scenarios can be realized by the mimetic $R^2$ and power-law $F(R)$ models. Using standard methods we calculate the observational indices in the slow-roll approximation and we show the compatibility to all the latest observational data can be achieved. In addition, we briefly address the graceful exit from inflation issue by studying the evolution of curvature perturbations when the slow-roll condition does not any longer holds true. Finally, the conclusions follow in the end of the paper.

\section{Unification of Inflation and Dark Energy Era with Mimetic $F(R)$ Gravity}

The mimetic $F(R)$ gravity formalism was introduced in \cite{Nojiri:2014zqa}, and the gravitational action is,
\be
\label{MF1}
S = \int \sqrt{-g} \left\{ \frac{F(R)}{2\kappa^2} + \lambda \left( \partial_\mu \phi \partial^\mu \phi
+ 1 \right) - V(\phi) \right\} + S_\mathrm{matter} \left( g_{\mu\nu}, \Psi \right)\, ,
\ee
where $S_\mathrm{matter}$ denotes the action of the matter fluids present.
We now rewrite the action (\ref{MF1}) in a scalar-tensor form by introducing
the auxiliary field $A$
\be
\label{MF1A}
S = \int \sqrt{-g} \left\{ \frac{1}{2\kappa^2} \left( F'(A) \left( R - A \right)
+ F(A) \right)
+ \lambda \left( \partial_\mu \phi \partial^\mu \phi
+ 1 \right) - V(\phi) \right\} + S_\mathrm{matter} \left( g_{\mu\nu}, \Psi \right)\, ,
\ee
The variation of the action (\ref{MF1A}) with respect to $A$ give the
equation $A=R$ and by substituting the obtained equation $A=R$ into
(\ref{MF1A}), we reobtain the action (\ref{MF1}).
By rescaling the metric by
\be
\label{JGRG22}
g_{\mu\nu}\to \e^\sigma g_{\mu\nu}\, ,\quad \sigma = -\ln F'(A)\, ,
\ee
we obtain the action in the Einstein frame,
\begin{align}
\label{MG2}
S =& \int \sqrt{ -g } \left\{ \frac{1}{2\kappa^2}
\left( R - \frac{3}{2} \partial_\mu \sigma
\partial^\mu \sigma - U\left( \sigma \right) \right) + \lambda \e^\sigma
\left( \partial_\mu \phi \partial^\mu \phi + \e^\sigma
\right) - \e^{2\sigma} V(\phi) \right\}
+ S_\mathrm{matter} \left( \e^\sigma g_{\mu\nu}, \Psi \right)\, , \nn
U\left( \sigma \right) =& \frac{A}{F'\left(A\left(\sigma\right)\right)}
 - \frac{F \left(A\left(\sigma\right)\right)}{F'\left(
 A\left(\sigma\right)\right)^2}\, .
\end{align}
Here $A(\sigma)$ is given by solving the equation
$\sigma =- \ln F'(A)$ with respect to $A$ as $A=A\left(\sigma\right)$.
By varying the resulting action with respect to the physical metric
$g_{\mu\nu}$, we obtain the following Einstein equations,
\begin{align}
\label{MG3}
0 =& g_{\mu\nu} \left\{
\frac{1}{2 \kappa^2} \left( R - \frac{3}{2} \partial_\mu \sigma
\partial^\mu \sigma - U\left( \sigma \right) \right) + \lambda \e^\sigma
\left( \partial_\mu \phi \partial^\mu \phi + \e^\sigma
\right) - \e^{2\sigma} V(\phi) \right\} \nn
& - \frac{1}{\kappa^2} \left( R_{\mu\nu} - \frac{3}{2} \partial_\mu \sigma
\partial_\nu \sigma \right) + \lambda \e^\sigma \partial_\mu \phi \partial_\nu \phi
+ T_{\mu\nu} \, ,
\end{align}
with $T_{\mu\nu}$ denoting the energy-momentum tensor corresponding to the matter fluids present.
By varying with respect to $\sigma$ yields,
\be
\label{MG4}
0 = \frac{1}{2\kappa^2}\left( 3 \nabla^\mu \nabla_\mu \sigma - U'(\sigma) \right)
+ \lambda \e^\sigma \left( \partial_\mu \phi \partial^\mu \phi + 2 \e^\sigma
\right) - 2 \e^{2\sigma} V(\phi) + 2 T\, ,
\ee
with $T\equiv T_\mu^{\ \mu}$. On the other hand, by the varying the action with respect to $\phi$, we obtain,
\be
\label{MG5}
0 = - 2 \nabla^\mu \left( \lambda \e^\sigma \nabla_\mu \phi \right) - \e^{2\sigma} V'(\phi)\, ,
\ee
and also by varying with respect to $\lambda$, we obtain the mimetic constraint,
\be
\label{MG6}
0=\partial_\mu \phi \partial^\mu \phi + \e^\sigma\, .
\ee
We now assume that the flat background can be a solution to the gravitational equations if we neglect the contribution from the matter fluids, so $T_{\mu\nu}=0$. Then we should require that $V(\phi)=0$ and there exists a value $\sigma_0$ of $\sigma$, with the property that, $U(\sigma_0) = U'(\sigma_0) =0$. Then the flat background solution is given by,
\be
\label{MG7}
g_{\mu\nu} = \eta_{\mu\nu}\, , \quad \sigma = \sigma_0 \, , \quad
\phi = \phi_0 + \e^{\frac{\sigma_0}{2}} t\, , \quad \lambda=0 \, ,
\ee
where $\phi_0$ is an arbitrary constant. Let us briefly investigate how the Newton law becomes in the case of mimetic $F(R)$ gravity. In order to investigate the Newton law, we consider the following perturbation of the metric,
\be
\label{MG8}
g_{\mu\nu} = \eta_{\mu\nu} + h_{\mu\nu} \, , \quad \sigma = \sigma_0 + \delta \sigma
\, , \quad
\phi = \phi_0 + \e^{\frac{\sigma_0}{2}} t + \delta \phi \, , \quad \lambda=\delta \lambda \, .
\ee
Consequently, owing to the fact that we assumed $U(\sigma_0) = U'(\sigma_0) =0$, we find $U(\sigma) = \mathcal{O} \left( {\delta \sigma}^2 \right)$.
Then by neglecting the terms $\mathcal{O} \left( {\delta \sigma}^3 \right)$ and we only keep the term of ${\delta \sigma}^2$ as follows,
\be
\label{MG9}
U(\sigma) \sim \frac{3}{2} m^2 { \delta \sigma}^2\, .
\ee
Here we have chosen the constant of the proportion in ${\delta \sigma}^2$
term as $\frac{3}{2} m^2$.
By comparing (\ref{MG9}) with the kinetic term of $\sigma$ in (\ref{MG3}),
we find that $m$ corresponds to the mass of the scalar field $\sigma$.
By linearizing Eq.~(\ref{MG3}), we obtain,
\be
\label{MG10}
\partial_\lambda \partial^\lambda h_{\mu\nu} - \partial_\mu \partial^\lambda
h_{\lambda\nu}
 - \partial_\nu \partial^\lambda h_{\lambda\mu} + \partial_\mu \partial_\nu h
+ \eta_{\mu\nu} \partial^\rho \partial^\sigma h_{\rho\sigma} - \eta_{\mu\nu}
\partial_\lambda \partial^\lambda h
= \kappa^2 \left( \e^{2\sigma_0} \delta_\mu^{\ t} \delta_\nu^{\ t}
\delta \lambda + T_{\mu\nu} \right)\, .
\ee
Here $h$ is the trace of $h_{\mu\nu}$ defined by
$h \equiv \eta^{\mu\nu} h_{\mu\nu}$.
Furthermore Eqs.~(\ref{MG4}), (\ref{MG5}), (\ref{MG6}) are linearized as follows,
\begin{align}
\label{MG11}
0 =& \frac{3}{2\kappa^2}\left( \partial^\mu \partial_\mu \delta \sigma
 - m^2 \delta \sigma \right)
+ \e^{2\sigma_0} \delta \lambda + 2 T\, , \\
\label{MG12}
0 =& \delta \dot\lambda \, , \\
\label{MG13}
0=& \delta \dot \phi \, .
\end{align}
Eqs.~(\ref{MG12}) and (\ref{MG13}) indicate that $\delta \lambda$ and $\delta\phi$ are the functions of the spatial coordinates $\bm{r}$. If $\delta \lambda=0$, Eq.~(\ref{MG10}) gives the standard Newton law, and Eq.~(\ref{MG11}) gives a correction to the Newton law which is identical to the standard $F(R)$ gravity correction (see \cite{Nojiri:2010wj} for example).
However, if $\delta \lambda (\bm{r}) \neq 0$, the term including $\delta \lambda (\bm{r})$ plays the role of the inhomogeneous dark matter and there could appear uncontrollable corrections to the Newton law. The correction is, however, time-independent, therefore if we choose the initial condition as follows, $\delta\lambda = 0$, no correction to the Newton law is generated. In the action (\ref{MF1}) in the Jordan frame, by varying with respect to the metric, instead of (\ref{MG3}) we obtain,
\begin{align}
\label{MG14}
0= & \frac{1}{2\kappa^2} \left\{ \frac{1}{2} g_{\mu\nu} F(R) - R_{\mu\nu} F' (R)
+ \nabla_\mu \nabla_\nu F'(R) - g_{\mu\nu} \nabla^2 F'(R) \right\} \nn
& + \frac{1}{2} g_{\mu\nu} \left\{ \lambda \left( \partial_\mu \phi \partial^\mu \phi + 1 \right)
 - V(\phi) \right\} - \lambda \partial_\mu \phi \partial_\nu \phi +\frac{1}{2} T_{\mu\nu} \, .
\end{align}
On the other hand, instead of (\ref{MG5}) and (\ref{MG6}), the variations of the action with respect to $\phi$ and $\lambda$ yield,
\begin{align}
\label{MG15}
0 =& - 2 \nabla^\mu \left( \lambda \nabla_\mu \phi \right) - V'(\phi)\, , \\
\label{MG16}
0 =& \partial_\mu \phi \partial^\mu \phi + 1\, .
\end{align}
We now assume that the background metric is a flat Friedmann-Robertson-Walker metric (FRW),
\be
\label{MG17}
ds^2 = - dt^2 + a(t)^2 \sum_{i=1,2,3} \left( dx^i \right)^2 \, ,
\ee
and $\phi$ only depends on $t$. Then Eq.~(\ref{MG16}) can be solved as follows,
\be
\label{MG18}
\phi = t \, ,
\ee
and effectively, Eq.~(\ref{MG15}) takes the following form,
\be
\label{MG19}
0 = 2 \dot \lambda + 6 H \lambda - V'(t)\, ,
\ee
where $H= \dot a /a$. On the other hand, the $(t,t)$ and $(i,j)$ $\left( i,j=1,2,3 \right)$ components of (\ref{MG14}) have the following form:
\begin{align}
\label{MG20}
0= & \frac{1}{2\kappa^2} \left\{ - \frac{1}{2} F(R) + 3 \left( H^2 + \dot H \right) F'(R)
 - 18 \left( 4 H^2 \dot H+ H \ddot H \right) F''(R) \right\} + \frac{1}{2} V(t)
 - \lambda + \frac{1}{2} \rho \, , \\
\label{MG21}
0= & \frac{1}{2\kappa^2} \left\{ \frac{1}{2} F(R) - \left( \dot H + 3 H^2 \right) F'(R)
+ 6 \left( 8 H^2 \dot H + 4 {\dot H}^2 + 6 H \ddot H + \dddot H \right) F''(R)
+ 36 \left( 4 H \dot H + \ddot H \right)^2 F''' (R) \right\} \nn
& - \frac{1}{2} V(t) + \frac{1}{2} p \, ,
\end{align}
where $\rho$ and $p$ are energy density and the pressure of the matter, respectively and the scalar curvature is given by $R=12H^2 + 6 \dot H$. By using the conservation law for the matter fluids, $0 = \dot \rho + 3 H \left( \rho + p \right)$, Eq.~(\ref{MG19}) can be obtained from (\ref{MG20}) and (\ref{MG21}). We should also note that Eq.~(\ref{MG20}) is the equation which governs the cosmic time $t$ dependence of the Lagrange multiplier $\lambda$, and it has the following form,
\begin{align}
\label{lambda}
\lambda(t) =& \frac{1}{2\kappa^2} \left\{ - \frac{1}{2} F \left( R\left(t\right) \right)
+ 3 \left( H(t)^2 + \dot H(t) \right) F' \left( R\left(t\right) \right)
 - 18 \left( 4 H(t)^2 \dot H(t) + H(t) \ddot H(t) \right) F'' \left( \left(R\right) \right) \right\} \nn
& + \frac{1}{2} V(t) + \frac{1}{2} \rho(t) \, .
\end{align}
Then the only remaining equation, which we should solve, is Eq.~(\ref{MG21}). Eq.~(\ref{MG21}) indicates that an arbitrary evolution of the expansion of the universe given by $a(t)$ can be realized by properly choosing the mimetic potential $V(\phi)$. If we assume that the energy density is given by the sum of the contributions from all the perfect fluids present, with constant Equation of State (EoS) parameter $w_i$, then the energy density $\rho_i$ with $w_i$ is proportional to $a^{-3(1+w_i)}$. Effectively, we find the following expression of $\rho$,
\be
\label{rho}
\rho = \sum_i \rho_{0i} a^{-3(1+w_i)}\, ,
\ee
with $\rho_{0i}$'s being constants. Eq.~(\ref{rho}) also gives the expression of the pressure $p$,
\be
\label{pressure}
p = \sum_i w_i \rho_{0i} a^{-3(1+w_i)}\, ,
\ee
and when $\rho$ and $p$ are given by (\ref{rho}) and (\ref{pressure}), for the evolution of the expansion given by $a(t)$, Eq.~(\ref{MG21}) gives the mimetic potential $V$ as a function of $t$, which has the following form,
\begin{align}
\label{Vt}
V(t) = & \frac{1}{\kappa^2} \left\{ \frac{1}{2} F\left( 12H(t)^2 + 6 \dot H(t) \right)
 - \left( \dot H(t) + 3 H(t)^2 \right) F'\left( 12H(t)^2 + 6 \dot H(t) \right) \right. \nn
& + 6 \left( 8 H(t)^2 \dot H(t) + 4 {\dot H(t)}^2 + 6 H(t) \ddot H(t) + \dddot H(t) \right)
 F''\left( 12H(t)^2 + 6 \dot H(t) \right) \nn
& \left. + 36 \left( 4 H(t) \dot H(t) + \ddot H(t) \right)^2 F''' \left( 12H(t)^2 + 6 \dot H(t) \right) \right\}
+ p \left( a \left( t \right) \right) \, .
\end{align}
We may choose the $F(R)$ gravity to describe a realistic model, where the accelerated expansion of the Universe occurs at late times and the Newton law can be reproduced by the Chameleon mechanism \cite{Khoury:2003rn}. Then we find the potential $V(\phi)$ which vanishes at late times. Even if we choose $F(R)$ as that of a ``realistic'' model, the expansion of the Universe becomes smooth due to the potential $V(\phi)$. The ``realistic'' $F(R)$ model includes a parameter $m\sim 10^{-33}\, \mathrm{eV}$ which has mass dimensions. The parameter $m$ plays the role of the effective cosmological constant. When the curvature $R$ is large enough compared with $m^2$, $R\gg m^2$, the $F(R)$ gravity behaves as follows,
\be
\label{HS1}
F(R) = R - c_1 m^2 + \frac{c_2 m^{2n+2}}{R^n}
+ \mathcal{O}\left(R^{-2n}\right)\, ,
\ee
where $c_1$, $c_2$, and $n$ are positive dimensionless constants. We now consider the model where $H$ is given by,
\be
\label{MG22}
H = \frac{H_I \e^{- \frac{t}{t_0}}}{\e^{-\frac{t}{t_0}} + 1} + H_L \, .
\ee
We may choose the parameters $H_I$ and $H_L$ as $H_I\gg H_L>0$ and $t_0>0$ is a constant. In the early-time regime, which corresponds to the limit $t\to -\infty$, the Hubble rate behaves as $H\to H_I$ and in the late-time regime $t\to +\infty$, it behaves as $H\to H_L$. Then the early-time behavior could correspond to the inflationary regime and the late-time behavior to the accelerating expansion of the present Universe. The Hubble rate $H$ in (\ref{MG22}) tells that the scale factor $a(t)$ is given by,
\be
\label{MG23}
a(t) = a_0 \left( \e^{-\frac{t}{t_0}} + 1 \right)^{- t_0 H_I} \e^{H_L t}\, ,
\ee
with $a_0$ being a constant. We may assume that the matter fluids are actually perfect fluids with constant equation of state (EoS) parameters $w_i$'s, so Eqs.~(\ref{rho}) and (\ref{pressure}) are modified accordingly. Then by using the expression (\ref{MG23}), we can find the $t$-dependence of $\rho$ and $p$, which tells the $t$-dependence, that is, $\phi$-dependence of the potential $V(\phi)$ by using (\ref{MG21}). Especially when we choose,
\be
\label{MG25}
6 H_L^2 = c_1 m^2 \, ,
\ee
the mimetic potential $V$ vanishes in the late time era, when $R\gg m^2$. In addition, Eq.~(\ref{MG22}) indicates that,
\be
\label{MG26}
\dot H = -\frac{H_I \e^{t/t_0}}{t_0 \left(\e^{t/t_0}+1\right)^2}\, ,
\ee
and the effective EoS parameter $w_\mathrm{eff}$ is given by,
\be
\label{MG27}
w_\mathrm{eff} = -1 +\frac{2 H_I \e^{t/t_0}}{3 t_0 \left(H_I+H_L \e^{t/t_0}+H_L\right)^2}\, ,
\ee
which is a smooth function of the cosmic time $t$. For the model (\ref{MG22}), we have,
\begin{align}
\label{MG27B}
\ddot H =& \frac{H_I \e^{t/t_0} \left(\e^{t/t_0}-1\right)}{t_0^2 \left(\e^{t/t_0}+1\right)^3}\, , \\
\label{MG27C}
\dddot H =& -\frac{H_I \e^{t/t_0} \left(-4 \e^{t/t_0}+\e^{\frac{2 t}{t_0}}+1\right)}{t_0^3 \left(\e^{t/t_0}+1\right)^4}\, .
\end{align}
Then the potential $V(\phi)$ in (\ref{Vt}) can be easily  found by replacing $\dot{H}$, $\ddot{H}$ and $\dddot{H}$ from the above expressions, but we omit it for brevity.

A more realistic model in comparison to (\ref{MG22}) could be given by
\be
\label{MG29}
H=H_L \coth \left( \frac{3}{4} H_L t \left( 1 + \frac{\frac{t}{t_0}}{\sqrt{ 1 + \frac{t^2}{t_0^2} }} \right)
+ \frac{H_L}{H_I} \right) \, .
\ee
We again assume $H_L \ll H_I$ and when $t\gg t_0$, the Hubble rate goes to that of the $\Lambda$CDM model, that is,
\be
\label{MG30}
H \to H_L \coth \left( \frac{3}{2} H_L t \right)\, .
\ee
On the other hand, when $t\ll - t_0$, $H$ goes to a constant, that is,
\be
\label{MG31}
H \to H_I \, ,
\ee
which may correspond to the inflation in the early universe. The effective EoS parameter $w_\mathrm{eff}$ is again a smooth function of $t$. In fact, owing to the fact that,
\be
\label{MG31B}
\dot H = \frac{3}{4}{H_L^2} \frac{ 1 + \frac{\frac{2t}{t_0} + \frac{t^3}{t_0^3}}{
\left( 1 + \frac{\frac{t}{t_0}}{\sqrt{ 1 + \frac{t^2}{t_0^2} }} \right)^{\frac{3}{2}}}}
{\sinh^2 \left( \frac{3}{4} H_L t \left( 1 + \frac{\frac{t}{t_0}}{\sqrt{ 1 + \frac{t^2}{t_0^2} }} \right) \right)} \, ,
\ee
we find
\be
\label{MG31C}
w_\mathrm{eff} = -1 + \frac{ 1 + \frac{\frac{2t}{t_0} + \frac{t^3}{t_0^3}}{
\left( 1 + \frac{\frac{t}{t_0}}{\sqrt{ 1 + \frac{t^2}{t_0^2} }} \right)^{\frac{3}{2}}} }
{2 \cosh^2 \left( \frac{3}{4} H_L t \left( 1 + \frac{\frac{t}{t_0}}{\sqrt{ 1 + \frac{t^2}{t_0^2} }} \right)\right) } \, .
\ee
If we choose $H_L$ as in Eq.~(\ref{MG25}), even for the model in (\ref{MG29}), the potential $V(\phi)$ becomes very small in the late-time regime, but if $R\gg m^2$, the expansion of the Universe is generated by the $F(R)$ and the energy density coming from $\phi$ must be very small. For the model (\ref{MG29}), if matter dominates, the potential $V(\phi)$ is small even for the matter domination epoch. We should also note that Eq.~(\ref{MG16}) tells $\partial_\mu \phi$ is always time-like, which means that $\phi$ is monotonically increasing function of $t$. Then even if some fluctuations of matter fluids occur, if $V(\phi)$ was small enough before the fluctuations, $V(\phi)$ remains small after the fluctuations. Therefore, the mimetic potential $V(\phi)$ does not give any corrections to the Newton law in the late-time regime and even if we consider the matter instability in \cite{Dolgov:2003px}, the situation is never changed from the standard $F(R)$ gravity. The role of the potential $V(\phi)$ is to connect the inflation epoch and the matter dominant epoch smoothly in the early universe and the model is not changed from the standard $F(R)$ gravity in the late-time regime, at least when certain aspects of the evolution are considered. As long as the value of the Hubble rate $H$ in the late time is given by (\ref{MG25}), the above situation is general for all the $F(R)$ gravities which satisfy the asymptotic behavior as in (\ref{HS1}), as in \cite{Hu:2007nk},
\be
\label{HS1}
F(R)= R -\frac{m^2 c_1 \left(R/m^2\right)^n}{c_2 \left(R/m^2\right)^n + 1}\, .
\ee
The same applies for exponential $F(R)$ gravities \cite{Elizalde:2011ds,Bamba:2010ws,Linder:2009jz,Cognola:2007zu,Oikonomou:2014gsa,Oikonomou:2013rba,Cognola:2016gjy} of the form,
\begin{equation}\label{expmodnocurv}
F(R)=R-2\Lambda \left ( 1- \e^{\frac{R}{\beta \Lambda}} \right )\, ,
\end{equation}
where $\Lambda$ is the cosmological constant corresponding to present time and the parameter $\beta$ is a positive free parameter of the order $\mathcal{O}(1)$.

\section{Inflationary Solutions and Compatibility with Observational Data}

The mimetic $F(R)$ framework offers much freedom for realizing various inflationary scenarios in a successful and viable way, with the viability of each model being determined on whether the model is compatible with observational data. In this section we shall demonstrate how to construct viable mimetic $F(R)$ inflation models, for various $F(R)$ gravities, and we shall demonstrate that the resulting models can be compatible with the observational data. In some recent works \cite{Odintsov:2015wwp}, the same issue has been addressed by using the perfect fluid approach, however in this paper we shall calculate the slow-roll indices and the corresponding observational indices by treating the mimetic theory as a particular case of an $F(R,\phi)$ scalar-tensor theory. The calculation will be done by assuming a slow-roll era, for which the following relations hold true,
\begin{equation}
\label{slowrollerarealtions}
\dot{H}\ll H^2\, ,\quad \ddot{H}\ll H\dot{H}\, ,
\end{equation}
where $H$ is the Hubble rate, and therefore the resulting gravitational equations become simplified as we show. The vacuum mimetic $F(R)$ gravity action of Eq.~(\ref{MF1}) can be viewed as a $F(R,\phi)$ scalar-tensor theory which takes the following form,
\be
\label{MF1sc}
S = \int \sqrt{-g} \left\{ \frac{F(R)}{2\kappa^2} + \lambda (\phi) \partial_\mu \phi \partial^\mu \phi
+ \lambda(\phi ) - V(\phi) \right\} \, ,
\ee
so effectively the kinetic term is $\omega (\phi)=-2 \lambda (\phi )$ and the potential is $\mathcal{V}(\phi)=\lambda (\phi)-V(\phi)$. The slow-roll indices of a general scalar-tensor theory of the form $F(R,\phi)$ were introduced in \cite{Noh:2001ia,Hwang:2001qk,Noh:2004rt,Hwang:2001pu} and have the following general form,
\begin{equation}
\label{slowrollindices}
\epsilon_1=-\frac{\dot{H}}{H^2}\, ,\quad \epsilon_2=\frac{\ddot{\phi}}{H\dot{\phi}}\, , \quad \epsilon_3=\frac{\dot{F'}(R,\phi)}{2HF'(R,\phi)}\, , \quad \epsilon_4=\frac{\dot{E}}{2HE}\, ,
\end{equation}
with the function $E(R,\phi)$ being equal to,
\begin{equation}
\label{epsilonfunction}
E(R,\phi)=F(R,\phi)\omega (\phi)+\frac{3\dot{F'}(R,\phi)^2}{2\dot{\phi}^2}\, .
\end{equation}
For the mimetic $F(R)$ gravity at hand, since $\phi=t$ and $\omega (\phi)=-2\lambda (\phi)$, the slow-roll indices become,
\begin{equation}
\label{slowrollindicesmim}
\epsilon_1=-\frac{\dot{H}}{H^2}\, ,\quad \epsilon_2=0\, ,\quad \epsilon_3=\frac{\dot{F'}(R,\phi)}{2HF'(R,\phi)}\, ,\quad \epsilon_4=\frac{\dot{E}}{2HE}\, ,
\end{equation}
with $E$ having the following form,
\begin{equation}
\label{epsilonfunction1}
E(R,\phi)=2F(R,\phi)\lambda (\phi)+\frac{3\dot{F'}(R,\phi)^2}{2}\, .
\end{equation}
In principle, for a given $F(R)$ gravity, the calculation of the slow-roll indices can be very tedious without using the slow-roll approximation, so in the following we shall use the slow-roll approximation and we will calculate the slow-roll indices and the corresponding observational indices of inflation. With regards to the latter, we are interested in the spectral index of primordial curvature perturbations $n_s$ and also to the scalar-to-tensor ratio $r$. In the slow-roll approximation, these two inflationary indices are defined in terms of the slow-roll indices $\epsilon_i$, $i=1,..,4$, as follows \cite{Noh:2001ia,Hwang:2001qk,Noh:2004rt,Hwang:2001pu},
\begin{equation}
\label{obsindices}
n_s\simeq 1-4\epsilon_1-2\epsilon_2+2\epsilon_3-2\epsilon_4\, ,\quad r=16(\epsilon_1+\epsilon_3)\, ,
\end{equation}
so in the following by specifying the $F(R)$ gravity and using the slow-roll approximation, we shall realize certain cosmological scenarios which are in concordance with the Planck 2015 data \cite{Ade:2015lrj,Planck:2013jfk}, which constrain $n_s$ and $r$ as follows,
\begin{equation}
\label{planckdata}
n_s=0.9644\pm 0.0049\, , \quad r<0.10\, .
\end{equation}
Also, the most recent BICEP2/Keck Array data \cite{Array:2015xqh}, constrain the scalar-to-tensor ratio as follows,
\begin{equation}
\label{scalartotensorbicep2}
r<0.07\, ,
\end{equation}
at $95\%$ confidence level, and as we demonstrate this constraint will be also satisfied by some of our models. We shall study two $F(R)$ gravity models, the $R^2$ inflation model, with the corresponding $F(R)$ gravity being,
\begin{equation}
\label{rsquare}
F(R)=R+\frac{R^2}{6M^2}\, ,
\end{equation}
and also the power-law model,
\begin{equation}
\label{frpowerlaw}
F(R)=\alpha R^n\, ,
\end{equation}
with $n>0$. The non-mimetic $R^2$ model is compatible with the Planck data, however the non-mimetic power-law model (\ref{rsquare}) is not in full concordance with observations, since $r$ is too large, but as we show, the mimetic versions of these two $F(R)$ models can be compatible with the Planck data and also with the BICEP2/Keck Array data.

\subsection{The Mimetic $R^2$ Model}

Assume that the $F(R)$ gravity is the $R^2$ model appearing in Eq.~(\ref{rsquare}), and by using the mimetic $F(R)$ formalism we shall realize the following cosmological evolution,
\begin{equation}
\label{hubblersquare}
H(t)=H_0-\frac{d}{6}(t-t_0)+c (t-t_0)^2\, ,
\end{equation}
where $H_0$, $d$, $c$ and $t_0$ are arbitrary parameters, which have to obey the following relation in order for the slow-roll approximation to hold true,
\begin{equation}
\label{relationsrquare}
H_0>d,c,t_0\, .
\end{equation}
Also $t_0\ll 1$, since it is the time instance at which the inflationary era begins. In addition, the slow-roll approximation holds true for the Hubble rate (\ref{hubblersquare}) since the cosmic time variables takes values in the interval $(10^{-35},10^{-15})$, and effectively the time dependent terms in Eq.~(\ref{hubblersquare}) are extremely small. It is conceivable that in order for the mimetic $F(R)$ theory to realize the cosmology (\ref{hubblersquare}), the mimetic potential and the Lagrange multiplier have to be appropriately chosen. In order to find their analytic form, we shall simplify the gravitational equations of motion which correspond to the action (\ref{MF1sc}) by using the slow-roll approximation. Particularly, since the relations (\ref{slowrollerarealtions}) hold true, the $F'(R)$ gravity is simplified as $F'(R)\simeq R/(3M^2)$ and also $\dot{F'}\simeq 8 H \dot{H}/M^2$, so the gravitational equations for the mimetic $R^2$ gravity read,
\begin{equation}
\label{rsquareprofinal}
\ddot{H}-\frac{\dot{H}^2}{2 H}+\frac{M^2H}{2}=-3H\dot{H}+M^2\frac{V(\phi)-2\lambda (\phi)}{12 H}\, ,
\end{equation}
so by neglecting the first two terms we finally get,
\begin{equation}
\label{rsquarefinal}
\frac{M^2H}{2}=-3H\dot{H}+M^2\frac{V(\phi)-2\lambda (\phi)}{12 H}\, .
\end{equation}
Then, by using the cosmological evolution of Eq.~(\ref{hubblersquare}), we can find in a straightforward way the potential
and the Lagrange multiplier which can generate the cosmological evolution, and the result is,
\begin{equation}\label{vlambdacombo}
V(t)-2\lambda (t)=\frac{\left(-d+M^2+12 c (t-t_0)\right) (-6 H_0+(t-t_0) (d+6 c (-t+t_0)))^2}{6 M^2}\, .
\end{equation}
The analytic forms of the potential and of the Lagrange multiplier can be found by using equation (\ref{MG19}) and combining it with Eq.~(\ref{vlambdacombo}), so the Lagrange multiplier reads,
\begin{equation}
\label{lambdap}
\lambda (\phi)=4 c \phi -\frac{10 c d \phi }{M^2}-\frac{120 c^2 t_0 \phi }{M^2}
+\frac{60 c^2 \phi ^2}{M^2}+\mathcal{A}_1\, ,
\end{equation}
where the constant $\mathcal{A}_1$ is,
\begin{equation}
\label{constanta}
\mathcal{A}_1=-\frac{d}{3}+\frac{d^2}{3 M^2}
+\frac{12 c H_0}{M^2}-4 c t_0+\frac{10 c d t_0}{M^2}+\frac{60 c^2 t_0^2}{M^2}\, .
\end{equation}
Correspondingly, the mimetic potential can easily be found by using (\ref{vlambdacombo}), and the explicit form appears in the Appendix. In the slow-roll approximation, the slow-roll indices $\epsilon_3$ and $\epsilon_4$ become approximately equal to,
\begin{equation}
\label{slowrollapprox}
\epsilon_3\simeq -2\epsilon_1\, ,\quad \epsilon_4\simeq \frac{\dot{\lambda}}{2H\lambda}\, ,
\end{equation}
while $\epsilon_1$ remains as it was. By using the expressions for $\lambda (t)$ and $H(t)$ given in Eqs.~(\ref{lambdap}) and (\ref{hubblersquare}) respectively, the observational indices take a very simple form,
\begin{align}
\label{obsfinalrsquare}
& \epsilon_1=\frac{\frac{d}{6}-2 c (t-t_0)}{\left(H_0-\frac{1}{6} d (t-t_0)+c (t-t_0)^2\right)^2}\, ,\quad
\epsilon_3\simeq -\frac{2 \left(\frac{d}{6}-2 c (t-t_0)\right)}{\left(H_0-\frac{1}{6} d (t-t_0)+c (t-t_0)^2\right)^2} \nn
& \epsilon_4\simeq \frac{18 c \left(-5 d+2 \left(M^2+30 c (t-t_0)\right)\right)}{\left(d^2-d \left(M^2
+30 c (t-t_0)\right)+12 c \left(3 H_0+\left(M^2+15 c (t-t_0)\right) (t-t_0)\right)\right) (6 H_0-(t-t_0) (d+6 c (-t+t_0)))}\, ,
\end{align}
and by using the $e$-foldings number $N$, the spectral index $n_s$ is found to be,
\begin{equation}
\label{spectralrsquare}
n_s\simeq 1-\frac{48 (d-12 c z)}{(-6 H_0+z (d-6 c z))^2}-\frac{36 c \left(-5 d+2 M^2+60 c z\right)}{(6 H_0+z (-d+6 c z)) \left(d^2-d \left(M^2+30 c z\right)+12 c \left(3 H_0+z \left(M^2+15 c z\right)\right)\right)}\, ,
\end{equation}
where the $N$-dependence of the spectral index is contained in the parameter $z$ which is given in the Appendix. Accordingly, the scalar-to-tensor ratio reads,
\begin{equation}
\label{sctotensorr}
r\simeq -\frac{16 \left(\frac{d}{6}-2 c z\right)}{\left(H_0-\frac{d z}{6}+c z^2\right)^2}\, .
\end{equation}
By using the following values for the parameters,
\begin{equation}
\label{valuesparmeters}
N=50\, ,\quad c=0.0000009\, ,\quad M=0.085 \, ,\quad H_0=0.26\, ,\quad d=0.002\, ,
\end{equation}
we obtain the following values for the observational indices,
\begin{equation}
\label{observindicesfinalresults}
n_s\simeq 0.967429\, ,\quad r\simeq 0.0134796\, ,
\end{equation}
which are compatible with both the Planck data of Eq.~(\ref{planckdata}) and with the BICEP2/Keck Array data of Eq.~(\ref{scalartotensorbicep2}). Thus even within the context of the mimetic gravity, the $R^2$ model is compatible with the recent observational data. However, in this case the utility of the mimetic framework is not shown, so in the next section we shall discuss the power law gravity of Eq.~(\ref{frpowerlaw}) which in the non-mimetic gravity framework is not compatible with the observational data. As we will show, it can be compatible with observations in the mimetic framework and this shows clearly the utility of the mimetic framework.

We need to note that the resulting picture we presented in this section crucially depends on the values of the parameter $H_0$, and in order to obtain viable results, certain fine-tuning is required. For example, if instead of choosing $H_0=0.26$, we choose, $0.24<H_0<0.29$, then the compatibility with the observational data cannot be achieved for any value of the rest of the parameters. A detailed analysis can show that only for the following values of the parameters, compatibility with the Planck data can be achieved:
\begin{equation}\label{valuesrange}
H_0=0.27\pm 0.03,\,\,\,M=0.08\pm 0.005,\,\,\,c/ d\sim 2\times 10^{3},\,\,\, \frac{H_0}{d}\sim 140\, ,
\end{equation}
for $N$ taking values in the interval $N=(50,60)$, and also we omitted the physical units for simplicity. Therefore, the model we just presented can be considered that it has restricted viability due to the required fine-tuning of the parameters, which leads to a very restricted allowed parameter space. Finally, this fine-tuning crucially depends on the form of the $F(R)$ gravity and correspondingly it depends on the form of the mimetic potential, and this is reasonable since the $F(R)$ gravity is fixed from the beginning of the calculation. In addition it crucially depends on the realized evolution, and this also justifies the fine-tuning for the specific model. This behavior is expected to occur in other models too, however as we show in the next section, the resulting behavior is extremely model dependent. So in principle there should be no explicit general rule for the amount of fine-tuning required.

\subsection{A Mimetic Power-law $F(R)$ Model}

One of the appealing features that the mimetic $F(R)$ framework offers is that it is possible to realize in a observationally viable way various cosmological scenarios for various $F(R)$ gravities. In addition, it is possible for an $F(R)$ gravity to produce a cosmologically viable inflationary era, even in the case that the non-mimetic $F(R)$ gravity could not achieve that. In this section we shall study an $F(R)$ gravity of this form, and particularly the power-law model of Eq.~(\ref{frpowerlaw}). In the ordinary $F(R)$ gravity case, the slow-roll indices of Eq.~(\ref{slowrollindices}) in the slow-roll regime, are equal to,
\begin{equation}
\label{slowrollfrordinary}
\epsilon_1=\frac{2-n}{(n-1)(2 n-1)}\, ,\quad \epsilon_2=0\, ,\quad \epsilon_3=\frac{(1-n) (2-n)}{(-1+n) (-1+2 n)}\, ,\quad \epsilon_4=\frac{n-2}{n-1}\, ,
\end{equation}
and the spectral index $n_s$ is therefore equal to,
\begin{equation}
\label{nsordfr}
n_s\simeq \frac{-7+5 n}{1-3 n+2 n^2}\, ,
\end{equation}
while the corresponding scalar-to-tensor ratio is,
\begin{equation}
\label{sttrationfrord}
r\simeq \frac{16 (-2+n)^2}{(-1+n) (-1+2 n)}\, .
\end{equation}
Since the variable $n$ is the only free parameter in the theory, only the value $n=1.8105$ makes the spectral index compatible with the Planck data (\ref{planckdata}), however the corresponding scalar-to-tensor ratio is not compatible with observations, as it can be seen below,
\begin{equation}
\label{obsrveatinindordfr}
n_s\simeq 0.966191\, ,\quad r\simeq 0.27047\, .
\end{equation}
Therefore, the ordinary power-law $F(R)$ gravity of Eq.~(\ref{frpowerlaw}) does not produce a viable inflationary era. However, the corresponding mimetic theory generates a viable inflationary era, as we now show.

Consider the following cosmological evolution,
\begin{equation}
\label{hubblefrpowerlaw}
H(t)=H_0-\frac{M^2}{6}(t-t_0)\, ,
\end{equation}
which is assumed to be a quasi de Sitter evolution, so the parameters must satisfy $H_0>M^2$. Regardless of the constraints obeyed by the parameters, the cosmic time $t$ takes very small values since the inflationary era occurs at times of the order $\sim 10^{-35}$\,sec, so the slow-roll conditions (\ref{slowrollerarealtions}) hold true for the Hubble rate (\ref{hubblefrpowerlaw}). We shall demonstrate that the cosmology (\ref{hubblefrpowerlaw}) can be realized by a mimetic theory and we shall show that this cosmology is compatible with both the Planck and the BICEP2/Keck Array data. Indeed, the gravitational equations in the slow-roll approximation become in this case,
\begin{equation}
\label{powerlawfrgravequations}
\alpha n=2\alpha (n-1)-\alpha n (n-1)24 \dot{H}+\frac{V(t)-2\lambda (t)}{6(12H^2)^{n-1}H^2}\, ,
\end{equation}
so the corresponding Lagrange multiplier has the following simple form,
\begin{equation}
\label{lagrafrsimple}
\lambda (\phi)=-2^{-3+2 n} 3^{-3+n} M^2 n \left(2-n+4 M^2 (-1+n) n\right) \alpha \left(H_0+\frac{1}{6} M^2 (t_0-\phi )\right)^{2 (-1+n)}\, ,
\end{equation}
and the mimetic potential reads in this case,
\begin{equation}
\label{mimeticpotfrpowerlaw}
V(\phi)=3^{-3+n} 4^{-1+n} \left(2-n+4 M^2 (-1+n) n\right) \alpha \left(9-\frac{36 M^2 n}{\left(6 H_0+M^2 (t_0-\phi )\right)^2}\right) \left(H_0+\frac{1}{6} M^2 (t_0-\phi )\right)^{2 n}\, .
\end{equation}
The slow-roll indices as functions of the cosmic time for the power-law $F(R)$ gravity (\ref{frpowerlaw}) in the slow-roll approximation are equal to,
\begin{equation}
\label{slowrollgeneralrelations}
\epsilon_3\simeq -(n-1)\epsilon_1\, ,\quad \epsilon_4\simeq \frac{\dot{\lambda}}{2H\lambda}\, ,
\end{equation}
and by combining Eqs.~(\ref{hubblefrpowerlaw}), (\ref{lagrafrsimple}) and (\ref{mimeticpotfrpowerlaw}) the resulting slow-roll indices as functions of the cosmic time take the following form,
\begin{equation}
\label{poorlone}
\epsilon_1=\frac{M^2}{6 \left(H_0-\frac{1}{6} M^2 (t-t_0)\right)^2}\, ,\quad \epsilon_3=\frac{M^2 (1-n)}{6 \left(H_0-\frac{1}{6} M^2 (t-t_0)\right)^2}\, ,\quad \epsilon_4=-\frac{6 M^2 (-1+n)}{\left(6 H_0+M^2 (-t+t_0)\right)^2}\, .
\end{equation}
It is more convenient to express the resulting observational indices as functions of the $e$-folding number, so the resulting expressions for the spectral index $n_s$ and for the scalar-to-tensor ratio are,
\begin{equation}
\label{obsrindifin}
n_s=\frac{3 H_0^2-M^2 (2+N)}{3 H_0^2-M^2 N}\, ,\quad r=\frac{8 M^2 (-2+n)}{-3 H_0^2+M^2 N}\, .
\end{equation}
By choosing the free variables and the $e$-foldings number as follows,
\begin{equation}
\label{powerlawfreeparametersvalues}
N=50\, ,\quad H_0=12.04566\, ,\quad M=2\, ,\quad n=1.8\, ,
\end{equation}
the observational indices take the following values,
\begin{equation}
\label{obsrvfinalvalues}
n_s\simeq 0.966\, ,\quad r\simeq 0.0272\, ,
\end{equation}
which are compatible to both the Planck and BICEP2/Keck Array data. More generally, if the parameters $M$ and $H_0$ are related as $H_0=6.0228\times M$, then by appropriately adjusting $n$, compatibility with the observational data can be achieved. Hence we showed that a viable quasi de Sitter cosmology can be realized by the mimetic power-law $F(R)$ model in a simple way. In principle, alternative scenarios to the one we used in this section may be generated. For example, consider the following cosmic evolution,
\begin{equation}
\label{newinfldesitter}
H(t)=H0 - d/6 (t - t_0) + c (t - t_0)^2\, ,
\end{equation}
and for the power-law $F(R)$ gravity (\ref{frpowerlaw}), the Lagrange multiplier that generates (\ref{newinfldesitter}) is at leading order,
\begin{equation}
\label{leadinglagra}
\lambda (\phi)\simeq -2^{1+2 n} 3^{-1+n} c H_0^{-1+2 n} (-1+n) n \alpha\, ,
\end{equation}
and the full form can be found in the Appendix. The corresponding mimetic potential can easily be found and it is approximately equal to,
\begin{equation}
\label{corrspotelagra}
V(\phi)= 12^{-1+n} H_0^{-1+2 n} \alpha \left(-16 c (-1+n) n+H_0 \left(2+4 n^2 (d+12 c (t_0-\phi ))-n (1+4 d+48 c (t_0-\phi ))\right)\right)\, .
\end{equation}
Correspondingly, the observational indices can be calculated but we do not quote it here for brevity. By using the following values for the parameters,
\begin{equation}
\label{finalvaluesforprm}
N=50\, ,\quad c=0.014\, ,\quad d=-0.1\, ,\quad n=11\, ,\quad H_0=11,
\end{equation}
the resulting values for the observational indices are,
\begin{equation}
\label{finalvaluesnsr}
n_s\simeq 0.966121\, ,\quad r\simeq 0.0176679\, ,
\end{equation}
which again are compatible with both the Planck and BICEP2/Keck Array data. Hence we showed that in the context of mimetic $F(R)$ gravity with potential and Lagrange multiplier, in principle it is possible to realize various cosmological scenarios in a successful way. What now remains is the graceful exit from inflation issue, which we address in the next section.

As we discussed in the previous section, for the $R^2$ model, a fine-tuning of the parameters was required in order to have compatibility with the observational data. However, as we show now, for the power-law model, the parameter space that may yield observationally viable results, is totally different from the one corresponding to the $R^2$ model. Particularly, it seems that the values of $n$, the exponent of the Ricci scalar in the $F(R)$ gravity function of Eq.~(\ref{frpowerlaw}), must take values approximately $n=1.7\pm 0.2$, and for this range of values for $n$, some numerical analysis can show that the compatibility is accomplished if the parameters satisfy the following relations,
\begin{equation}\label{restrictionsii}
\frac{H_0}{M}\sim 6.023\, ,
\end{equation}
for $N$ taking values in the interval $N=(50,60)$, and we omitted the physical units for simplicity. Obviously, in the present case too, certain fine-tuning is required in order to achieve compatibility, and this reduces the general viability of the model. Finally, we need to note that the fine-tuning is strongly model dependent, since the conditions for the viability of each model we studied, are different and are expected to be different in other models too. In the next section we shall also into account certain restrictions on the values of the parameter $n$, coming from the graceful exit issue.

\subsubsection{Graceful Exit for Power-law Mimetic $F(R)$ Gravity}

In this section we will discuss how the graceful exit from inflation can be achieved for the mimetic $F(R)$ gravity. For convenience we shall consider the power-law model of Eq.~(\ref{frpowerlaw}) and we shall use the potential and the Lagrange multiplier appearing in Eqs.~(\ref{lagrafrsimple}) and (\ref{mimeticpotfrpowerlaw}). Recall that the corresponding mimetic theory produces a viable cosmology for the values chosen as in Eq.~(\ref{powerlawfreeparametersvalues}), so for $n=1.8$, and as we will show the graceful exit from inflation strongly depends on the value of $n$. In order to study the graceful exit issue, we shall investigate the existence and the stability of de Sitter vacua of the mimetic theory, when the slow-roll approximation does not hold true anymore. Thus, the existence of an unstable de Sitter vacuum will generate growing curvature perturbations which will make the graceful exit from inflation possible. For more details on this mechanism consult Refs.~\cite{Odintsov:2015wwp,Bamba:2014jia} and references therein. Firstly, let us see how the gravitational equation (\ref{powerlawfrgravequations}) becomes in the case that the slow-roll approximation is abandoned. In order to extract the corresponding differential equation, we use Eq.~(\ref{MG20}) and also the $F(R)$ gravity of Eq.~(\ref{frpowerlaw}), so the corresponding gravitational equation of motion reads,
\begin{align}
\label{diffgrexit}
6 \alpha n H(t)^2-(\alpha n -\alpha )\left(12 H(t)^2+6H'(t)\right)
+6 \alpha n (n-1)(24 H(t) H'(t)+6 H''(t))-\frac{V(t)-2\lambda (t)}{12 H(t)^2+6H'(t)}=0\, .
\end{align}
Now we seek de Sitter vacua solutions for the equation (\ref{diffgrexit}), so we set $H(t)=H_d$, and the resulting equation is,
\begin{equation}
\label{gr1}
6 H_d^2 n \alpha -12 H_d^2 (-\alpha +n \alpha )-12^{-2+n} H_d^{-2+2 n} (-2 (-1+n) \alpha +n \alpha )=0\, ,
\end{equation}
which has a single solution,
\begin{equation}
\label{desittersolution}
H_d=\left(2^{-5+2 n} 3^{-3+n}\right)^{\frac{1}{4-2 n}}\, .
\end{equation}
In order to study the stability of this solution, we set $H(t)=H_d+\Delta H(t)$ in the differential equation (\ref{diffgrexit}) and we expand, keeping linear terms in $\Delta H(t)$, $\Delta H'(t)$ and $\Delta H''(t)$, so the resulting differential equation reads,
\begin{align}
\label{resultingdiffeqn}
& \frac{1}{288 H_d^2}\alpha \left(-4 H_d \left(864 H_d^2+12^n H_d^{2 n}\right) (-2+n) \Delta H(t)\right. \nn
& +\left(-12^n H_d^{2 n} (-2+n)-1728 H_d^2 (-1+n)+41472 H_d^3 (-1+n) n+3^{1+n} 4^{2+n} H_d^{1+2 n} (-1+n) n\right) \Delta H'(t) \nn
& \left.+2 H_d \left(\left(-864 H_d^2+12^n H_d^{2 n}\right) (-2+n)+5184 H_d (-1+n) n \Delta H''(t)\right)\right)=0\, .
\end{align}
By solving the above differential equation we obtain the following solution,
\begin{equation}
\label{desitterunstablesol}
\Delta H(t)=-\frac{864 H_d^2-2^{2 n} 3^n H_d^{2 n}}{2 \left(864 H_d^2+2^{2 n} 3^n H_d^{2 n}\right)}+c_1 \e^{\mu_1t}+c_2 \e^{\mu_2 t}\, ,
\end{equation}
where the parameters $\mu_1$ and $\mu_2$ appear in the Appendix and $c_1$, $c_2$ are arbitrary constants. By substituting the de Sitter solution (\ref{desittersolution}) and also by setting $n=1.8$ the solution (\ref{desitterunstablesol}) becomes,
\begin{equation}
\label{explicitsol1}
\Delta H(t)=-0.499989+c_1 \e^{0.0775432t}+c_2 \e^{0.00195451 t}\, ,
\end{equation}
which indicates that the perturbation $\Delta H(t)$ is fastly growing with time. Therefore, for the value of $n$ that makes the slow-roll approximated mimetic theory viable, we also make the non slow-rolling theory unstable, and therefore graceful exit is achieved via growing curvature perturbations. Before closing this section we need to note that a slight change in the parameter $n$, utterly changes the final picture. For example, by setting $n=1.5$, the resulting solution to equation (\ref{resultingdiffeqn}) is equal to,
\begin{equation}
\label{resultingstable1}
\Delta H(t)=-0.497691+c_1 \e^{-0.0409565t} \sin (0.0950736 t)\, ,
\end{equation}
where we have set $c_1=c_2$. Clearly, the solution (\ref{resultingstable1}) describes a decaying perturbation with time, so in effect the de Sitter solution (\ref{desittersolution}) is stable and therefore no graceful exit from inflation occurs. Thus it is intriguing that for the value of $n$ which renders the mimetic theory under study viable, we also achieve graceful exit from inflation.

Actually, it can be shown that for $n>1.6$, the curvature perturbations around the de Sitter solution are always growing. Therefore, combining this result with the fact the power-law model is compatible with observations for $n=1.7\pm 0.2$, then we can see that the two parameter spaces coincide for the interval $n=(1.6,1.9)$. Hence, in this interval the graceful exit from inflation can be qualitatively guaranteed, since the de Sitter vacua are unstable. However, as we already noted, this is a qualitative approach and does not really provides us with more concrete information with regards to the actual duration of inflation, and mainly how much do actually the perturbations grow for $N\sim 50-60$. In principle, the $e$-folding number can be easily reached, since the model contains many free parameters, but let us show this explicitly.

In the context of $F(R)$ gravity inflation, the graceful exit from inflation occurs when the first slow-roll parameter $\epsilon_1$ in Eq.~(\ref{slowrollindices}), becomes of order $\epsilon_1\sim \mathcal{O}(1)$, so let us assume that this occurs at the time instance $t=t_f$ at which point the Hubble rate is $H=H_f$. For the evolution under study, namely that of Eq.~(\ref{hubblefrpowerlaw}), the condition $\epsilon_1\sim 1$ yields $H_f^2\simeq \frac{M^2}{6}$, and in addition, at $t=t_f$, we have from Eq.~(\ref{hubblefrpowerlaw}),
\begin{equation}\label{hubblecrevis}
H_f=H_0-\frac{M^2}{6}(t_f-t_0)\, ,
\end{equation}
so by solving with respect to $t_f-t_0$ and substituting $H_f^2\simeq \frac{M^2}{6}$ we get,
\begin{equation}\label{fgrdsol}
t_f-t_0\simeq \frac{6}{M^2}\left (H_0-\frac{M}{\sqrt{6}} \right)\, .
\end{equation}
Then, by using the definition of the $e$-foldings number $N$ as a function of the Hubble rate, which is,
\begin{equation}\label{defrev}
N=\int_{t_0}^{t_f}H(t)\mathrm{d}t\, ,
\end{equation}
we can express the difference $(t_f-t_0)$ as a function of the $e$-foldings number $N$, so eventually we can have a quantitative numerical idea on how the perturbations of Eq.~(\ref{explicitsol1}), grow as a function of $N$. In order to be correct then, the initial time $t=t_0$ should be the time instance that the primordial curvature modes exit the horizon. By integrating (\ref{defrev}) for the Hubble rate (\ref{hubblefrpowerlaw}), we get,
\begin{equation}\label{nrev}
N=H_0(t_f-t_0)-\frac{M^2}{12}(t_f-t_0)^2\, ,
\end{equation}
so by substituting $(t_f-t_0)$ from Eq.~(\ref{fgrdsol}), we get,
\begin{equation}\label{finalalmostrev}
N\simeq \frac{3 H_0^2}{M^2}-\frac{1}{2}\, .
\end{equation}
Finally, by using Eqs.~(\ref{fgrdsol}) and (\ref{finalalmostrev}), we may express the quantity $(t_f-t_0)$ as a function of $N$ and the rest of the parameters, as follows,
\begin{equation}\label{dfrfefinal}
t_f-t_i\simeq \frac{2(N+\frac{1}{2})}{H_0}-\frac{6}{M\sqrt{6}}\, .
\end{equation}
Then, by substituting the quantity $(t_f-t_0)$, the perturbations of Eq.~(\ref{explicitsol1}) become\footnote{The time instance $t_0$ can appear in Eq.~(\ref{explicitsol1}), by appropriately choosing the free integration constants $c_1$ and $c_2$},
\begin{equation}
\label{explicitsol1new}
\Delta H(N)=-0.499989+c_1 \e^{0.0775432 \left(\frac{2(N+\frac{1}{2})}{H_0}-\frac{6}{M\sqrt{6}}\right)}+c_2 \e^{0.00195451\left(\frac{2(N+\frac{1}{2})}{H_0}-\frac{6}{M\sqrt{6}}\right)}\, ,
\end{equation}
and at this point we can have a concrete idea on how much the perturbations (\ref{explicitsol1new}) grow as
a function of $N,H_0,M$, for $n=1.8$. So for the values of $H_0$ and $M$ chosen as in
Eq.~(\ref{powerlawfreeparametersvalues}), the perturbations for $N=50$ become approximately,
\begin{equation}
\label{explicitsol1newrev1}
\Delta H(N)=-0.499989+c_1 \e^{0.0116629}+c_2 \e^{0.462714}\, ,
\end{equation}
which means that the perturbations are already of order $\mathcal{O}(1)$, even for $N=50$, since $\e^{0.462714}\sim 1.5$. However,  by inserting $H_0=12.04566$ and $M=2$, one
obtains,
\begin{align}
 \Delta
H(N)=&-0.499989+c_1\exp\left\{0.0775432\left[-\sqrt{\frac{3}{2}}+0.166035\left(\frac{1}{2}+N\right)\right]\right\} \nonumber \\
 &+c_2\exp\left\{0.00195451\left[-\sqrt{\frac{3}{2}}+0.166035\left(\frac{1}{2}+N
\right)\right]\right\}~. \nonumber
\end{align}
Therefore, even for $N$ small, i.e.~already at the beginning of inflation
($N\sim\mathcal{O}(1)$), one finds $\Delta H\sim -0.5+0.9c_1+1.0c_2$. Unless
the parameters $c_1$ and $c_2$ are chosen to be extremely small, one finds that
$\Delta H$ is already $\mathcal{O}(1)$ from the onset of inflation, hence the de
Sitter solution of the model at hand is highly unstable. In conclusion, inflation in the present model cannot last
until $N\gtrsim 50$, which is necessary to explain observations, and therefore fine-tuning is required in order to make the model a viable model of inflation.

\subsection{Late-time Evolution}

In this section we briefly study the late-time evolution of the mimetic power-law $F(R)$ model, for the cosmologies (\ref{hubblefrpowerlaw}) and (\ref{newinfldesitter}). We start off with the evolution (\ref{hubblefrpowerlaw}), so the total effective equation of state parameter becomes at late times,
\begin{equation}
\label{latetimeev1}
w_\mathrm{eff}\simeq -1+\frac{4}{M^2 t^2}\, ,
\end{equation}
which since $t\gg 1$, it clearly describes a nearly de Sitter (since the term second term is very small compared to $-1$) but slightly turned to quintessential evolution. So the late-time picture is that the power-law mimetic $F(R)$ model of Eq.~(\ref{frpowerlaw}) describes a quintessential late-time acceleration era. The corresponding potential $V(\phi)$ is equal to,
\begin{equation}
\label{pot1latetimes}
V(\phi)\simeq \mathcal{B}\phi^{2n}\, ,
\end{equation}
and correspondingly, the Lagrange multiplier is,
\begin{equation}
\label{lambda1latetimes}
\lambda (\phi)=\mathcal{A} \phi^{2 (n-1)}\, ,
\end{equation}
with $\mathcal{A}$ and $\mathcal{B}$ being constants which can be found in the Appendix. In the same way we can find the late-time behavior of the evolution (\ref{newinfldesitter}), in which case the effective equation of state parameter at very late times reads,
\begin{equation}
\label{secondeffatverylattetimes}
w_\mathrm{eff}\simeq -1-\frac{4}{3 c t^3}\, ,
\end{equation}
which since $t\gg 1$, describes a nearly de Sitter ($w_\mathrm{eff}\simeq -1$), but slightly turned to phantom (since $\frac{4}{3 c t^3}\ll 1$) acceleration era. Therefore, this case is different from the previous case, where we had a quintessential acceleration era. The corresponding potential at leading order reads,
\begin{equation}
\label{corrpotentlatetimes1}
V(\phi )\simeq -2^{1+4 n} 3^{-1+2 n} c^{1+4 n} n^2 \alpha ^2 \phi^{1+8 n}\, ,
\end{equation}
and the Lagrange multiplier is,
\begin{equation}
\label{laqgranewnswqa}
\lambda (\phi)\simeq -2^{1+2 n} 3^{-1+n} c^{2 n} n \left(-1-3 n+4 n^2\right) \alpha \phi^{-2+4 n}\, .
\end{equation}
Hence as we demonstrated, the mimetic power-law $F(R)$ gravity model leads to nearly quintessential or nearly phantom late-time acceleration, depending on the choice of the potential and of the corresponding Lagrange multiplier. In addition, we need to stress that the powers of the scalar field $\phi$, crucially control this late-time behavior.

\section{Conclusions}

In this paper we studied how the unification of late and early-time acceleration can be achieved in the context of mimetic $F(R)$ gravity. We also discussed how to realize various viable inflationary scenarios for two $F(R)$ gravity models, namely the $R^2$ model and a power-law model of the form $F(R)=\alpha R^n$. We calculated the slow-roll indices and the corresponding observational indices in the slow-roll approximation and we demonstrated that the mimetic models can be compatible with the latest observational data. Therefore, even though the power-law model in the context of non-mimetic $F(R)$ gravity was not in full agreement with the observational data, in the context of the mimetic theory the model is in full agreement with the data. In addition, the graceful exit for the power-law model is generated by growing curvature perturbations when the slow-roll approximation breaks down. Interestingly enough, the parameter $n$ crucially affects the behavior of the curvature perturbations, making these grow or decay, depending on the value of $n$. As we showed, for the value of $n$ that the mimetic power-law $F(R)$ model is compatible with the observational data, the curvature perturbations grow, thus the graceful exit from the inflationary era is guaranteed. However, the analysis showed that fine-tuning is needed in order to produce enough inflation by the time that the slow-roll era ends.

With this work we revealed some appealing aspects of mimetic $F(R)$ gravity and therefore one is confronted with the question: should mimetic gravity be considered as a mathematical construction or it captures certain features of a fundamental theory? In our opinion any cosmological theory which can be compatible with observations, in some sense effectively describes to some extent the fundamental theory which controls the Universe evolution. So it is a limit of the underlying theory which describes the Universe, and it deserves more analysis in order to reveal where it fails to describe the evolution. A drawback of the theory is that in some cases the potential and the Lagrange multiplier have complicated forms, but the advantage of the theory is that nothing is added by hand in the theory since the conformal degree of freedom of the metric is used only. Effectively, the scalar field naturally appears in
the theory in a geometrical way and admittedly after the Higgs discovery, it seems that the scalar fields are fundamental ingredients of
nature. So what remains is to reveal the actual interplay between the scalar field(s) and gravity, therefore the mimetic framework needs to further be explored.

\section*{Acknowledgments}

This work is supported (in part) by MEXT KAKENHI Grant-in-Aid for Scientific Research on Innovative Areas
 (No. 15H05890) (S.N.),
MINECO (Spain), project
 FIS2013-44881 (S.D.O), by the CSIC I-LINK1019 Project (S.D.O and S.N), and by Min. of Education and Science of Russia (S.D.O
and V.K.O).

\section*{Appendix: Explicit Forms of the Mimetic Potential and of Parameters}

In this Appendix we present the detailed forms of various parameters and functions appearing in the main text. The explicit form of the mimetic potential appearing in (\ref{vlambdacombo}) for the Lagrange multiplier chosen as in Eq.~(\ref{lambdap}), is equal to,
\begin{align}
\label{mimetcpot1}
V(\phi)=& -\frac{1}{6 M^2}\left(12 \left(-3 H_0^2 M^2+2 c \left(-3 H_0 \left(2+M^2 (t_0-\phi )^2\right)+18 H_0^2 (t_0-\phi )+2 M^2 (t_0-\phi )\right)\right.\right. \nn
& \left.-3 c^2 \left(20-24 H_0 (t_0-\phi )+M^2 (t_0-\phi )^2\right) (t_0-\phi )^2+36 c^3 (t_0-\phi )^5\right) \nn
& +d^3 (t_0-\phi )^2-d^2 \left(4-24 c t_0^3-12 H_0 (t_0-\phi )+M^2 (t_0-\phi )^2\right. \nn
& \left.+72 c t_0^2 \phi -72 c t_0 \phi ^2+24 c \phi ^3\right)+4 d \left(9 H_0^2+M^2 \left(1-3 c (t_0-\phi )^3\right)
\right. \nn
& \left.\left.+15 c \left(-2+3 c (t_0-\phi )^3\right) (t_0-\phi )-3 H_0 (t_0-\phi ) \left(M^2+18 c (-t_0+\phi )\right)\right)\right)\, .
\end{align}
The parameter $z=t-t_0$ appearing in Eq.~(\ref{spectralrsquare}) is equal to,
\begin{align}
\label{parameterz}
z=& \frac{M^2}{12 c}-\frac{144 c H_0-M^4}{6\ 2^{2/3} c \left(-432 c H_0 M^2+2 M^6+5184 c^2 N+\sqrt{4 \left(144 c H_0-M^4\right)^3+\left(-432 c H_0 M^2+2 M^6+5184 c^2 N\right)^2}\right)^{1/3}} \nn
& +\frac{\left(-432 c H_0 M^2+2 M^6+5184 c^2 N+\sqrt{4 \left(144 c H_0-M^4\right)^3+\left(-432 c H_0 M^2+2 M^6+5184 c^2 N\right)^2}\right)^{1/3}}{12\ 2^{1/3} c}\, .
\end{align}
The full form of the Lagrange multiplier appearing in Eq.~(\ref{leadinglagra}) is,
\begin{align}
\label{fullformlagra}
& \lambda (\phi)= 3^{-2+n} 4^{-1+n} n \left(H_0-\frac{1}{6} (t-t_0) (d+6 c (-t+t_0))\right)^{-1+2 n}\nn
& \times \left(-24 c (-1+n) \alpha +\frac{(d+12 c (-t+t_0)) \left(2-n (1+4 d-48 c t+48 c t_0)+4 n^2 (d+12 c (-t+t_0))\right) \alpha }{-6 H_0+(t-t_0) (d+6 c (-t+t_0))}\right)\, ,
\end{align}
and the corresponding potential appearing in Eq.~(\ref{corrspotelagra}) is,
\begin{align}
\label{corrpotlagraful}
V(\phi )=& 3^{-2+n} 4^{-1+n} \left(H_0-\frac{1}{6} (t-t_0) (d+6 c (-t+t_0))\right)^{2 n} \nn
& \times (3 (2-n+4 (-1+n) n (d+12 c (-t+t_0))) \alpha \nn
& \left.+\frac{2 n \left(-24 c (-1+n) \alpha +\frac{(d+12 c (-t+t_0)) \left(2-n (1+4 d-48 c t+48 c t_0)+4 n^2 (d+12 c (-t+t_0))\right) \alpha }{-6 H_0+(t-t_0) (d+6 c (-t+t_0))}\right)}{H_0-\frac{1}{6} (t-t_0) (d+6 c (-t+t_0))}\right)\, .
\end{align}
Moreover, the parameters $\mu_1$ and $\mu_2$ appearing in Eq.~(\ref{desitterunstablesol}) are equal to,
\be
\label{mu1andmu2}
\mu_1=\frac{q_1+\sqrt{q_2}}{2 \left(-10368 H_d^2 n+10368 H_d^2 n^2\right)}\, ,\quad \mu_2=\frac{q_1-\sqrt{q_2}}{2 \left(-10368 H_d^2 n+10368 H_d^2 n^2\right)}\, ,
\ee
where $q_1$ is,
\begin{align}
\label{q1}
q_1=& -1728 H_d^2-2^{1+2 n} 3^n H_d^{2 n}+1728 H_d^2 n+41472 H_d^3 n \nn
& +2^{2 n} 3^n H_d^{2 n} n+2^{4+2 n} 3^{1+n} H_d^{1+2 n} n-41472 H_d^3 n^2-2^{4+2 n}
3^{1+n} H_d^{1+2 n} n^2\, ,
\end{align}
and $q_2$ is,
\begin{align}
\label{q2}
q_2 = & -4 \left(6912 H_d^3+2^{3+2 n} 3^n H_d^{1+2 n}-3456 H_d^3 n-2^{2+2 n} 3^n H_d^{1+2 n} n\right) \left(-10368 H_d^2 n+10368 H_d^2 n^2\right) \nn
& +\left(1728 H_d^2+2^{1+2 n} 3^n H_d^{2 n}-1728 H_d^2 n-41472 H_d^3 n-2^{2 n} 3^n H_d^{2 n} n-2^{4+2 n} 3^{1+n} H_d^{1+2 n} n \right. \nn
& \left. +41472 H_d^3 n^2+2^{4+2 n} 3^{1+n} H_d^{1+2 n} n^2\right)^2 \, .
\end{align}
Also, the constant parameter $\mathcal{B}$ which appears in Eq.~(\ref{pot1latetimes}) is equal to,
\begin{equation}
\label{mathcalbpar}
\mathcal{B}=2^{-3+2 n} 3^{-2+n} M^{4 n} \left(2-n+4 M^2 (-1+n) n\right) \alpha\, ,
\end{equation}
and also the parameter $\mathcal{A}$ appearing in Eq.~(\ref{lambda1latetimes}) is,
\begin{equation}
\label{mathcalalatetimes}
\mathcal{A}= -2^{-3-2 (-1+n)+2 n} 3^{-3-2 (-1+n)+n} M^{2+4 (-1+n)} n \left(2-n+4 M^2 (-1+n) n\right) \alpha\, .
\end{equation}

\end{document}